\documentclass[rsi,reprint,superscriptaddress,nofootinbib]{revtex4-1}

\usepackage{graphicx}
\usepackage{subfigure}
\usepackage{amssymb}
\usepackage{amsmath}
\usepackage{multirow}
\usepackage{color}
\usepackage{comment}

\begin{document}

\title{A cryogenic rotation stage with a large clear aperture\\for the half-wave plates in the \textsc{Spider} instrument}

\author{Sean Bryan}
\affiliation{School of Earth and Space Exploration, Arizona State University, Tempe, AZ}
\email[Corresponding Author: ]{sean.a.bryan@asu.edu}

\author{Peter Ade}
\affiliation{School of Physics and Astronomy, Cardiff University, Cardiff, UK}

\author{Mandana Amiri}
\affiliation{Department of Physics and Astronomy, University of British Columbia, Vancouver, BC, Canada}

\author{Steven Benton}
\affiliation{Department of Physics, Princeton University, Princeton, NJ}
\affiliation{Department of Astronomy and Astrophysics, University of Toronto, Toronto, ON, Canada}

\author{Richard Bihary}
\affiliation{Department of Physics and CERCA, Case Western Reserve University, Cleveland, OH}

\author{James Bock}
\affiliation{Division of Physics, Mathematics $\&$ Astronomy, California Institute of Technology, Pasadena, CA}
\affiliation{Jet Propulsion Laboratory, Pasadena, CA}

\author{J. Richard Bond}
\affiliation{Canadian Institute for Theoretical Astrophysics, University of Toronto, Toronto, ON, Canada}

\author{H. Cynthia Chiang}
\affiliation{School of Mathematics, Statistics $\&$ Computer Science, University of KwaZulu-Natal, Durban, South Africa}

\author{Carlo Contaldi}
\affiliation{Theoretical Physics, Blackett Laboratory, Imperial College, London, UK}

\author{Brendan Crill}
\affiliation{Division of Physics, Mathematics $\&$ Astronomy, California Institute of Technology, Pasadena, CA}
\affiliation{Jet Propulsion Laboratory, Pasadena, CA}

\author{Olivier Dore}
\affiliation{Division of Physics, Mathematics $\&$ Astronomy, California Institute of Technology, Pasadena, CA}
\affiliation{Jet Propulsion Laboratory, Pasadena, CA}

\author{Benjamin Elder}
\affiliation{Department of Physics and CERCA, Case Western Reserve University, Cleveland, OH}

\author{Jeffrey Filippini}
\affiliation{Department of Physics, University of Illinois at Urbana-Champaign, IL}

\author{Aurelien Fraisse}
\affiliation{Department of Physics, Princeton University, Princeton, NJ}

\author{Anne Gambrel}
\affiliation{Department of Physics, Princeton University, Princeton, NJ}

\author{Natalie Gandilo}
\affiliation{Department of Physics, University of Toronto, Toronto, ON, Canada}

\author{Jon Gudmundsson}
\affiliation{Department of Physics, Princeton University, Princeton, NJ}

\author{Matthew Hasselfield}
\affiliation{Department of Astrophysical Sciences, Princeton University, Princeton, NJ}
\affiliation{Department of Physics and Astronomy, University of British Columbia, Vancouver, BC, Canada}

\author{Mark Halpern}
\affiliation{Department of Physics and Astronomy, University of British Columbia, Vancouver, BC, Canada}

\author{Gene Hilton}
\affiliation{National Institute of Standards and Technology, Boulder, CO}

\author{Warren Holmes}
\affiliation{Jet Propulsion Laboratory, Pasadena, CA}

\author{Viktor Hristov}
\affiliation{Division of Physics, Mathematics $\&$ Astronomy, California Institute of Technology, Pasadena, CA}

\author{Kent Irwin}
\affiliation{Department of Physics, Stanford University, Stanford, CA}

\author{William Jones}
\affiliation{Department of Physics, Princeton University, Princeton, NJ}

\author{Zigmund Kermish}
\affiliation{Department of Physics, Princeton University, Princeton, NJ}

\author{Craig Lawrie}
\affiliation{Department of Physics and CERCA, Case Western Reserve University, Cleveland, OH}

\author{Carrie MacTavish}
\affiliation{Kavli Institute for Cosmology, University of Cambridge, Cambridge, UK}

\author{Peter Mason}
\affiliation{Division of Physics, Mathematics $\&$ Astronomy, California Institute of Technology, Pasadena, CA}

\author{Krikor Megerian}
\affiliation{Jet Propulsion Laboratory, Pasadena, CA}

\author{Lorenzo Moncelsi}
\affiliation{Division of Physics, Mathematics $\&$ Astronomy, California Institute of Technology, Pasadena, CA}

\author{Thomas Montroy}
\affiliation{Department of Physics and CERCA, Case Western Reserve University, Cleveland, OH}

\author{Tracy Morford}
\affiliation{Division of Physics, Mathematics $\&$ Astronomy, California Institute of Technology, Pasadena, CA}

\author{Johanna Nagy}
\affiliation{Department of Physics and CERCA, Case Western Reserve University, Cleveland, OH}

\author{C. Barth Netterfield}
\affiliation{Department of Physics, University of Toronto, Toronto, ON, Canada}
\affiliation{Department of Astronomy and Astrophysics, University of Toronto, Toronto, ON, Canada}
\affiliation{Canadian Institute for Advanced Research CIFAR Program in Cosmology and Gravity, Toronto, ON, Canada}

\author{Ivan Padilla}
\affiliation{Department of Astronomy and Astrophysics, University of Toronto, Toronto, ON, Canada}

\author{Alexandra S. Rahlin}
\affiliation{Department of Physics, Princeton University, Princeton, NJ}

\author{Carl Reintsema}
\affiliation{National Institute of Standards and Technology, Boulder, CO}

\author{Daniel C. Riley}
\affiliation{Department of Physics and CERCA, Case Western Reserve University, Cleveland, OH}

\author{John Ruhl}
\affiliation{Department of Physics and CERCA, Case Western Reserve University, Cleveland, OH}

\author{Marcus Runyan}
\affiliation{Jet Propulsion Laboratory, Pasadena, CA}

\author{Benjamin Saliwanchik}
\affiliation{Department of Physics and CERCA, Case Western Reserve University, Cleveland, OH}

\author{Jamil Shariff}
\affiliation{Department of Physics and CERCA, Case Western Reserve University, Cleveland, OH}
\affiliation{Department of Astronomy and Astrophysics, University of Toronto, Toronto, ON, Canada}

\author{Juan Soler}
\affiliation{Institut d'Astrophysique Spatiale, Orsay, France}

\author{Amy Trangsrud}
\affiliation{Jet Propulsion Laboratory, Pasadena, CA}
\affiliation{Division of Physics, Mathematics $\&$ Astronomy, California Institute of Technology, Pasadena, CA}

\author{Carole Tucker}
\affiliation{School of Physics and Astronomy, Cardiff University, Cardiff, UK}

\author{Rebecca Tucker}
\affiliation{Division of Physics, Mathematics $\&$ Astronomy, California Institute of Technology, Pasadena, CA}

\author{Anthony Turner}
\affiliation{Jet Propulsion Laboratory, Pasadena, CA}

\author{Shyang Wen}
\affiliation{Department of Physics and CERCA, Case Western Reserve University, Cleveland, OH}

\author{Donald Wiebe}
\affiliation{Department of Physics and Astronomy, University of British Columbia, Vancouver, BC, Canada}

\author{Edward Young}
\affiliation{Department of Physics, Princeton University, Princeton, NJ}

\begin{abstract}
We describe the cryogenic half-wave plate rotation mechanisms built for and used in \textsc{Spider}, a polarization-sensitive balloon-borne telescope array that observed the Cosmic Microwave Background at 95 GHz and 150 GHz during a stratospheric balloon flight from Antarctica in January 2015. The mechanisms operate at liquid helium temperature in flight. A three-point contact design keeps the mechanical bearings relatively small but allows for a large (305 mm) diameter clear aperture. A worm gear driven by a cryogenic stepper motor allows for precise positioning and prevents undesired rotation when the motors are depowered. A custom-built optical encoder system monitors the bearing angle to an absolute accuracy of $\pm 0.1^{\circ}$. The system performed well in \textsc{Spider} during its successful 16 day flight.
\end{abstract}

\maketitle

\section{Introduction}

The ability to precisely rotate optical elements such as polarizers or wave plates is often crucial for the operation of polarimeters, and can provide a useful cross check for systematic errors. Millimeter-wave systems often need to be cooled to cryogenic temperatures to increase their sensitivity. Precise rotation of large, cryogenic parts poses a suite of challenges.  These include the limited cooling power available at cryogenic temperatures, the risks to vacuum integrity and the system complexity associated with mechanical shaft feedthroughs, and the difficulty of managing differential thermal contractions in parts with tight mechanical tolerances.

The \textsc{Spider} \cite{rahlin14,fraisse11} instrument is a balloon-borne array of six telescopes that imaged the polarization of the Cosmic Microwave Background at 95 GHz and 150 GHz during a 16 day flight in January 2015, with the goal of constraining or detecting a signature of cosmic inflation \cite{baumann09}. To reduce radiative loading on the detectors, which operate at 250~mK, the lenses and half-wave plate (HWP) polarization modulators are cooled with liquid helium to 4 K.

To enable strong control of beam systematics \cite{mactavish08,odea11}, each HWP in \textsc{Spider} is mounted skyward of its telescope's primary optic, which means that the rotation mechanisms need a large clear aperture of 305 mm. Simulations \cite{fraisse11} show that to achieve \textsc{Spider}'s science, each rotation mechanism needs to be able to rotate its HWP to any desired angle with an absolute accuracy of $\pm 1.0^{\circ}$. So, we set a target of $\pm 0.1^{\circ}$ precision to be well within that goal. The HWPs are cooled to 4 K to reduce in-band emission and to ensure that any reflections from the HWP terminate on cold surfaces instead of on a warm surface outside the cryostat, both of which reduce in-band optical loading on the bolometric detectors.  Since \textsc{Spider} uses the HWPs in a step-and-integrate mode, the mechanisms must prevent the HWPs from rotating while the instrument is observing. To facilitate lab characterization and reduce downtime during flight, we require that the mechanisms must turn smoothly at a minimum of 1$^\circ$ per second while cold. The rotation mechanisms must, on average, generate only a small amount of heat to conserve liquid helium during flight. Boiling off $\sim75$ L of liquid helium would have shortened the flight by an entire day \cite{gudmundsson15}, so we required that the mechanisms dissipate far less than that amount.

\begin{figure}
\begin{center}
\includegraphics[width=0.48\textwidth]{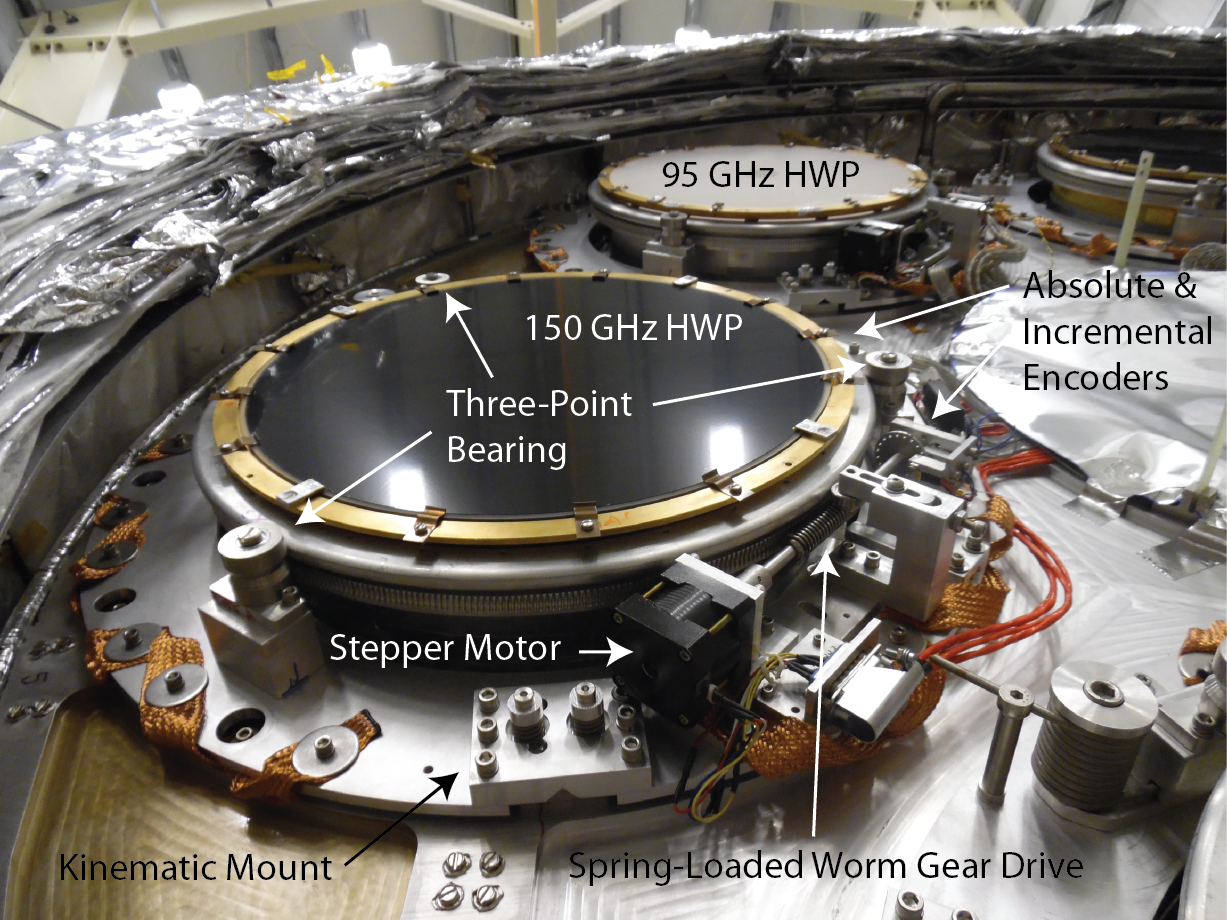}
\caption{Photograph of the HWP rotation mechanisms installed in the \textsc{Spider} flight cryostat. Each HWP stack is held in a 305 mm clear diameter Invar mounting ring attached to a rotor wheel and the main gear. The rotor wheel is held in a three-point bearing, while the main gear is driven by a cryogenic stepper motor  turning a worm shaft. The worm shaft is spring loaded toward the main gear, but is held approximately 0.010" from direct contact by an idler wheel contacting the rotor wheel. Optical encoders verify that main gear is at the desired angle. A kinematic mount prevents slight thermal deformations in the cryostat from mechanically distorting the mechanism. 
\label{hwp_mechanism_overview}}
\end{center}
\end{figure}

To meet these performance requirements, we used a worm gear configuration driven by a cryogenic stepper motor.  The large-diameter HWP rotor is held in a three-point mechanical bearing. The three-point bearing provides low-friction and relative immunity from thermal contractions, while the worm gear prevents the bearing from moving more than $0.05^{\circ}$ even when the stepper motor is not energized. We use custom-built cryogenic optical encoders to determine the bearing angle. A photograph of a rotation mechanism is shown in Figure~\ref{hwp_mechanism_overview}.

In this paper, we describe the engineering of the rotation mechanisms we built for \textsc{Spider}. We also present results from lab testing of the mechanical and cryogenic performance, and briefly discuss the successful operation of the mechanisms in flight.

\section{Comparison with Other Instruments}

\subsection{Optical/Instrument Approaches}

Several successful CMB polarization experiments have not used polarization modulation hardware, such as SPTpol \cite{austermann12}, BOOMERanG \cite{montroy06,jones07} and several seasons of ACTPol \cite{henderson15}. The sky appears to rotate for instruments at locations other than the South Pole site, which provides some degree of polarization modulation without any extra hardware in the instrument. This enabled BOOMERanG and the earlier seasons of ACTPol to successfully make CMB polarization maps in both polarization states with each detector in the instrument.  Others CMB instruments have used several different polarization modulation techniques, and here we present a brief overview of some of the effective approaches in the field today.

Several instruments, including the BICEP/Keck series of instruments \cite{b2iii15} and QUaD \cite{hinderks09}, have modulated their polarization sensitivity by rotating the entire instrument about the telescope boresight. This approach has the advantage of only requiring moving parts outside the cryogenic space. A disadvantage is that while the polarization sensitivity rotates, beam and instrumental systematic effects also rotate. Those systematic effects can be seperated from the true sky signal in data analysis \cite{b2iii15}. So far, the mechanical components required for boresight rotation have been too large and heavy for use on balloon missions.

The Maxipol \cite{johnson06} and EBEX experiments \cite{didier15} use continuously-rotating HWPs, and BICEP1 used Faraday Rotation Modules \cite{moyerman13} for one season of observations. These modulators were located near the focal plane of the instrument. This season of ACTPol, as well as the upcoming Advanced ACT instrument, also uses a continuously-rotating HWP near the focal plane \cite{henderson15}. Continuous rotation has the ability to modulate the sky signal up to an audio frequency above the detector 1/f knee. On the other hand, continuous modulation limits the maximum scan speed across the sky (the instrument can scan no faster than a fraction of a beam per modulation cycle time). A slower scan could move the signal closer to the detector 1/f knee, and could limit the ability of the instrument to simultaneously map CMB temperature anisotropies.

Like boresight roation, placing the modulator near the focal plane has the disadvantage of modulating the telescope's instrumental polarization and beam systematics along with the true sky signal. The ABS team \cite{kusaka14} addressed this by placing their continuously-rotating HWP at ambient temperature skyward of the entire telescope, so it only modulates the sky signal and not instrument systematics. However, differential loss and differential reflection lead to modulation-synchronous detector signals that they needed to remove in their data analysis.

A complimentary technology to HWPs is the Variable-delay Polarization Modulator, such as the one used in the CLASS instrument \cite{tomeh14}. This device consists of a wire grid suspended a distance above a flat mirror. Modulation is accomplished by either varying the mirror-grid distance, or fixing the distance and rotating the entire device. Understanding the device performance requires physical optics modeling \cite{chuss11}, which is similar to the situation with other modulators. The device is fabricated from conventional materials, and can operate at ambient temperature, but suspending the wire grid becomes technologically challenging for large diameters. In its variable-distance mode, the device modulates both linear and circular polarization, which is a disadvantage when observing sources such as the CMB which are almost totally linearly-polarized.

For \textsc{Spider}, we installed a HWP at liquid helium temperature skyward of each telescope. This means it does not modulate beam or instrumental systematics, enabling those to be more easily separated from sky signals. Since \textsc{Spider} scans rapidly to modulate the sky signal well above the detector 1/f knee, we did not need to use continuous polarization modulation. Instead, we stepped the HWP to a different angle twice each day. Since the HWP was not moving during observations, unlike the continuous-modulation systems, differential loss and differential transmission did not inject any spurious signals into the detectors. We selected a set of HWP angles that will allow us to combine maps to measure the true polarization on the sky, and isolate the effects of beam systematics, HWP differential transmission and HWP modulation efficiency \cite{bryanAO,bryanthesis}.

\subsection{Mechanical Approaches}

Our motor solution differs from the methods used by some other experiments in the field. The torque for the HWP rotation mechanisms in Maxipol \cite{johnson06} , EBEX \cite{didier15},  BLASTPol \cite{fissel10}, POLARBEAR \cite{kermish12} and PILOT \cite{salatino10} is provided via a rotating shaft fed through the vacuum wall of the cryostat. Since \textsc{Spider} is an array of six telescopes in a single cryostat, having six independent rotating shaft feed-throughs is undesirable. We therefore chose to use cryogenic stepper motors. This means that the only connections outside the cryostat for a single mechanism are four high-current ($\sim1$ A) wires, and 14 low current angle encoder readout wires 

To measure the rotation angle of our HWP bearing, \textsc{Spider} uses optical encoders mounted inside the cryostat. This also differs from the approach taken in other instruments. The HWP encoder in BLASTPol consists of a leaf spring making contact with a potentiometer located near the bearing. Maxipol, EBEX, and PILOT all have optical encoders. Maxipol has a commercial absolute optical encoder outside the cryostat. EBEX has a relative optical encoder on the main bearing. PILOT has a 3 bit absolute encoder on its main bearing which allows the bearing to step between 8 discrete angles. In contrast with \textsc{Spider} and EBEX which use optical encoders located at cryogenic temperature close to the HWP, PILOT has optical fibers running from the cryogenic HWP to the inner wall of the vacuum vessel. This allows for ambient temperature light sources and detectors.

\section{Rotation Stage Hardware}
\subsection{HWP Mount}

Each HWP in \textsc{Spider} consists of a 330 mm diameter birefringent single-crystal A-cut sapphire\footnote{GT Crystal Systems, Salem, MA www.gtat.com} slab with an anti-reflection (AR) coating. The thickness of the sapphire  is 3.18 mm for the 150 GHz band, and 4.93 mm for the 95 GHz band. For the 95 GHz HWP, we apply a 427~$\mu$m-thick fused quartz\footnote{Mark Optics, Santa Ana, CA www.markoptics.com} AR coat to each side of the sapphire.  These AR coats are bonded to the center of the sapphire with a small drop of Eccobond 24 adhesive, leaving a negligible $<$10~$\mu$m air gap. For the 150 GHz HWPs, we bonded a 254~$\mu$m-thick Cirlex AR coating layer onto each side of the sapphire.  This bond was made by melting a thin polyethylene layer between the Cirlex and sapphire, under pressure in an oven.  These AR coatings were tested over multiple thermal cycles to 77 K, as well as several cycles to 4 K, prior to flight. More details of the fabrication of the HWPs are presented in \cite{bryanthesis}

\begin{figure}
\begin{center}
\includegraphics[width=0.48\textwidth]{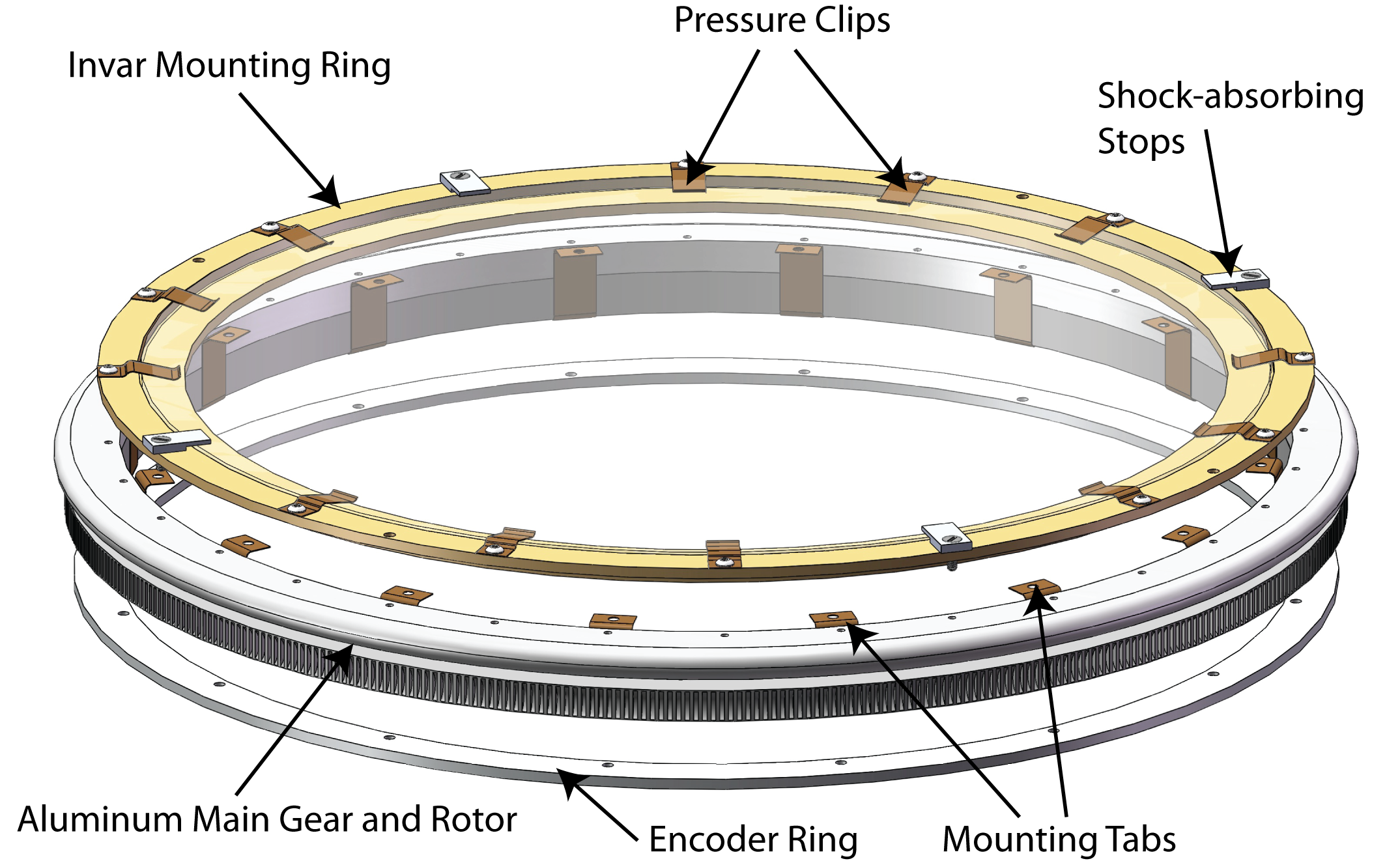}
\caption{Exploded view of an Invar HWP mount and aluminum main gear. The HWP stack is held down by 12 pressure clips and 4 shock-absorbing stops to ensure that the HWP does not come out of the mount in the event of a mechanical shock such as can be present when the payload is launched, or terminated. Sixteen radially compliant mounting tabs allow for differential thermal contraction between the main gear and HWP mount. \label{exploded_hwp_mount}}
\end{center}
\end{figure}

Each HWP is mounted in its rotation mechanism as shown in Figure~\ref{exploded_hwp_mount}. The mount needs to hold the HWP securely to prevent movement during shocks that could be experienced during shipping, launch, and flight termination. To achieve this, we use 12 phosphor bronze clips to apply a total of 16 lbs (i.e. 4 times the weight force of the HWP itself) of pressure to hold the HWP in an Invar mounting plate. As added protection against shock, we added 4 aluminum tabs wrapped with enough Kapton tape to almost touch the HWP. These limiting stops provide a cushion of Kapton tape that will stop the HWP if a hard upward shock is applied to it, instead of allowing that shock to deform the phosphor bronze mounting clips. To prevent corrosion, the Invar was plated\footnote{GCG Corporation, Glendale, CA (818) 247-8508} with 25 micro-inches of gold. Invar was used to reduce the impact of differential thermal contraction between the mounting ring and the HWP system. When an aluminum ring was used, which has much higher differential thermal contraction, the fragile quartz AR coating wafers would occasionally crack or shatter catastrophically. The Invar/clip system solved this problem.

The Invar-mounted HWP is attached to the aluminum main gear\footnote{GoTo \& Tracking Systems, Arlington, TX www.gototelescopes.com} and bearing with 16 flexible phosphor bronze mounting tabs. The tabs are rigid enough in the lateral direction to mount the HWP securely, but flexible enough radially to repeatedly comply with the differential thermal contraction between aluminum and Invar.

\subsection{Bearings}

To hold the HWP in place yet allow it to rotate smoothly, we use three V-groove guide bearings evenly spaced around the circumference of the rotor. This allows for a large clear aperture, but keeps the individual bearings small to minimize the effect of thermal contraction on the moving bearing parts. Each of the three guide bearings consists of a stainless steel V-groove turning on two stainless steel\footnote{SR4Z-Kit8526 VXB, Anaheim, CA www.vxb.com} ball bearings.  In the stepper motor described in Section~\ref{steppermotor}, a similar\footnote{S625Z-Kit8525} ball bearing is used. The ball bearings are shipped with a light oil coating for lubrication and to prevent corrosion, but the oil coating will seize up the bearing at cryogenic temperatures. To remove the oil, we cleaned the ball bearings with acetone and isopropyl alcohol. We chose the lowest performance grade of ball bearing, ABEC-1, because it has the largest mechanical clearances available, which is an advantage when faced with thermal contraction. To take up thermal contraction as the aluminum main gear shrinks, we spring-loaded one of the guide bearings with a torsion spring\footnote{9287K136 McMaster-Carr, Elmhurst, IL www.mcmaster.com} to push it radially inwards as the mechanism cools. A hard stop was attached to almost touch the spring-loaded part to make it impossible for the rotor to spring out of the mount. Since the rotor is thermally connected to the cold plate only via three point contacts, it was necessary to verify that it is sufficiently heat sunk to the 4 K mounting surface. We measured that in the temperature range from 10 K to 30 K, the thermal conductance between the main gear and the cold plate is sufficient, approximately 2 mW/K. Measurements of the liquid helium boiloff rate in the flight cryostat suggest that in flight each HWP absorbed 0.5 mW of thermal radiation or less. This means the HWPs were likely less than 0.5 mW / 0.2 mW/K = 2.5 K above liquid helium temperature.

\subsection{Motor}
\label{steppermotor}

\begin{figure}
\begin{center}
\includegraphics[width=0.48\textwidth]{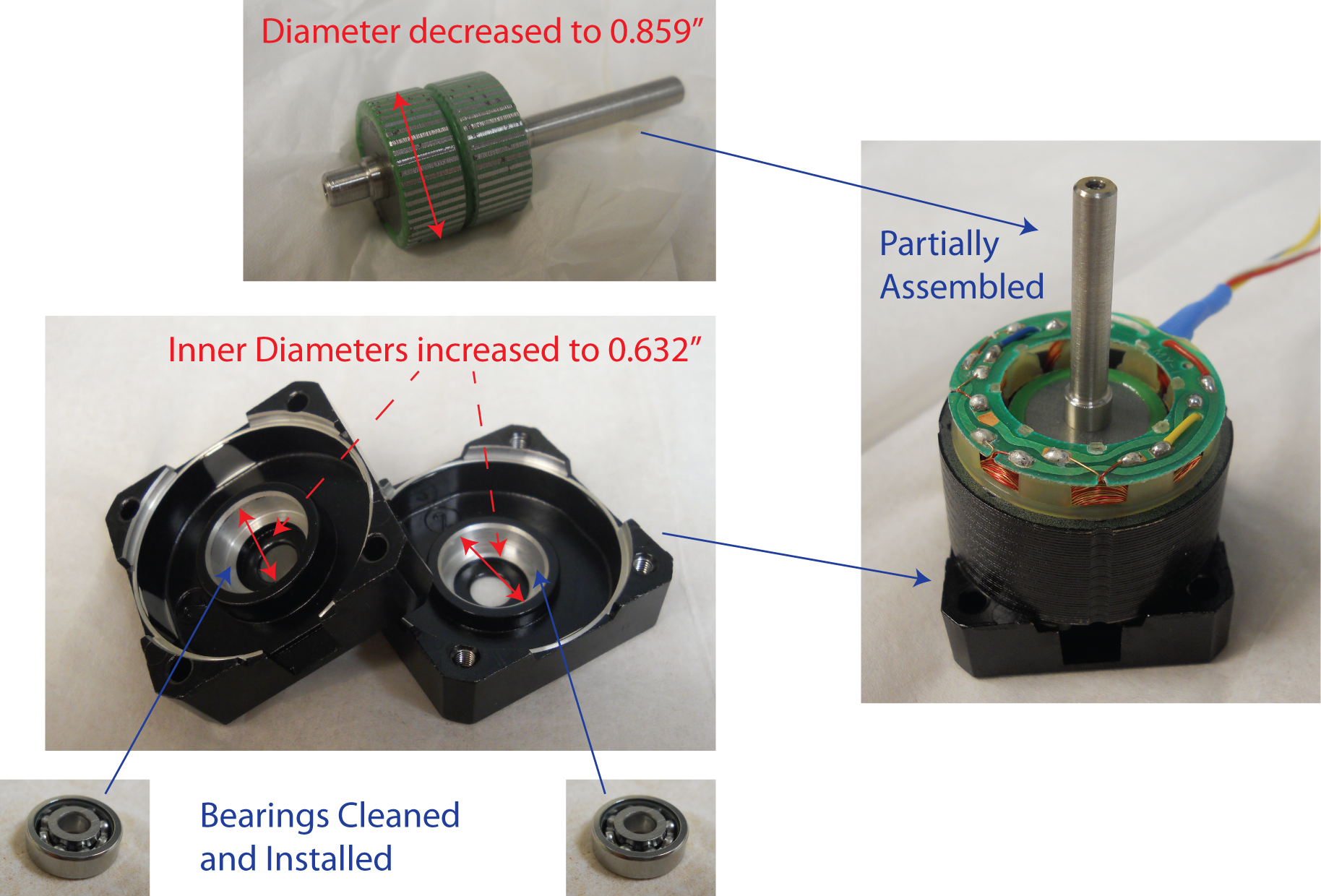}
\caption{Modifications to a Mycom PS445-01A stepper motor for cryogenic operation. To accommodate differential thermal contraction, we increased the clearance between the rotor (shown in the top left) and stator, as well as the clearance between the rotor bearings (bottom left) and the motor housing (center left). The right panel shows a partially-assembled motor. The bottom half of the motor housing, the stator, and the rotor are visible. \label{motor_mods}}
\end{center}
\end{figure}

We use a stepper motor\footnote{PS445-01A Mycom, Kyoto, Japan www.mycom-japan.co.jp} to provide the torque for the rotation mechanism. As long as the motor torque is sufficient this allows us to set the angle of the rotor precisely by the motor driver without the need for realtime feedback from the angle encoders. The motor is rated at 1.2 A drive current; we nominally drive at 0.8 A, and a minimum of 0.1-0.2 A is needed to overcome friction in the rest of the rotation mechanism when cold.

\begin{figure*}[ht]
\begin{center}
\includegraphics[width=0.49\textwidth]{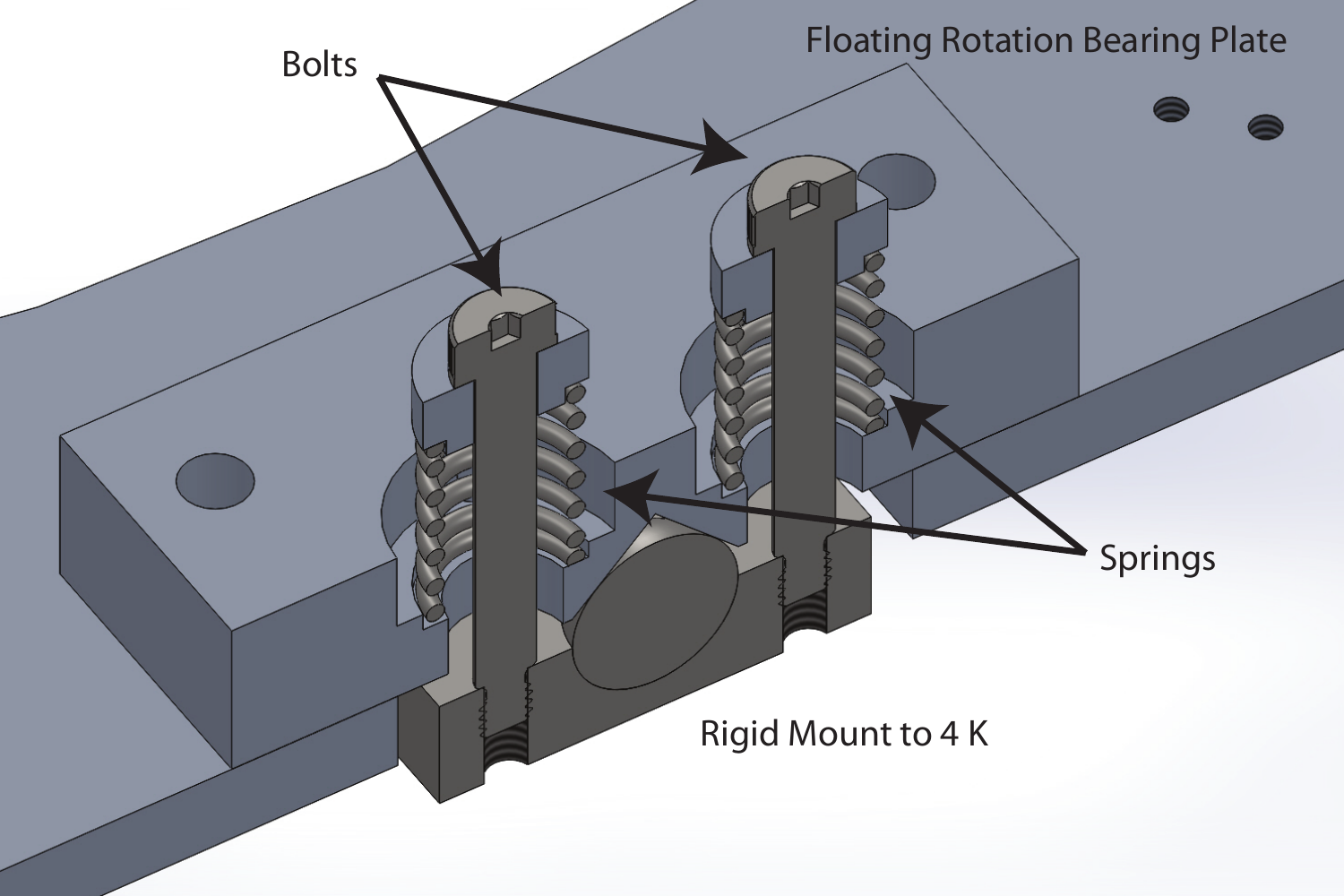}
\includegraphics[width=0.49\textwidth]{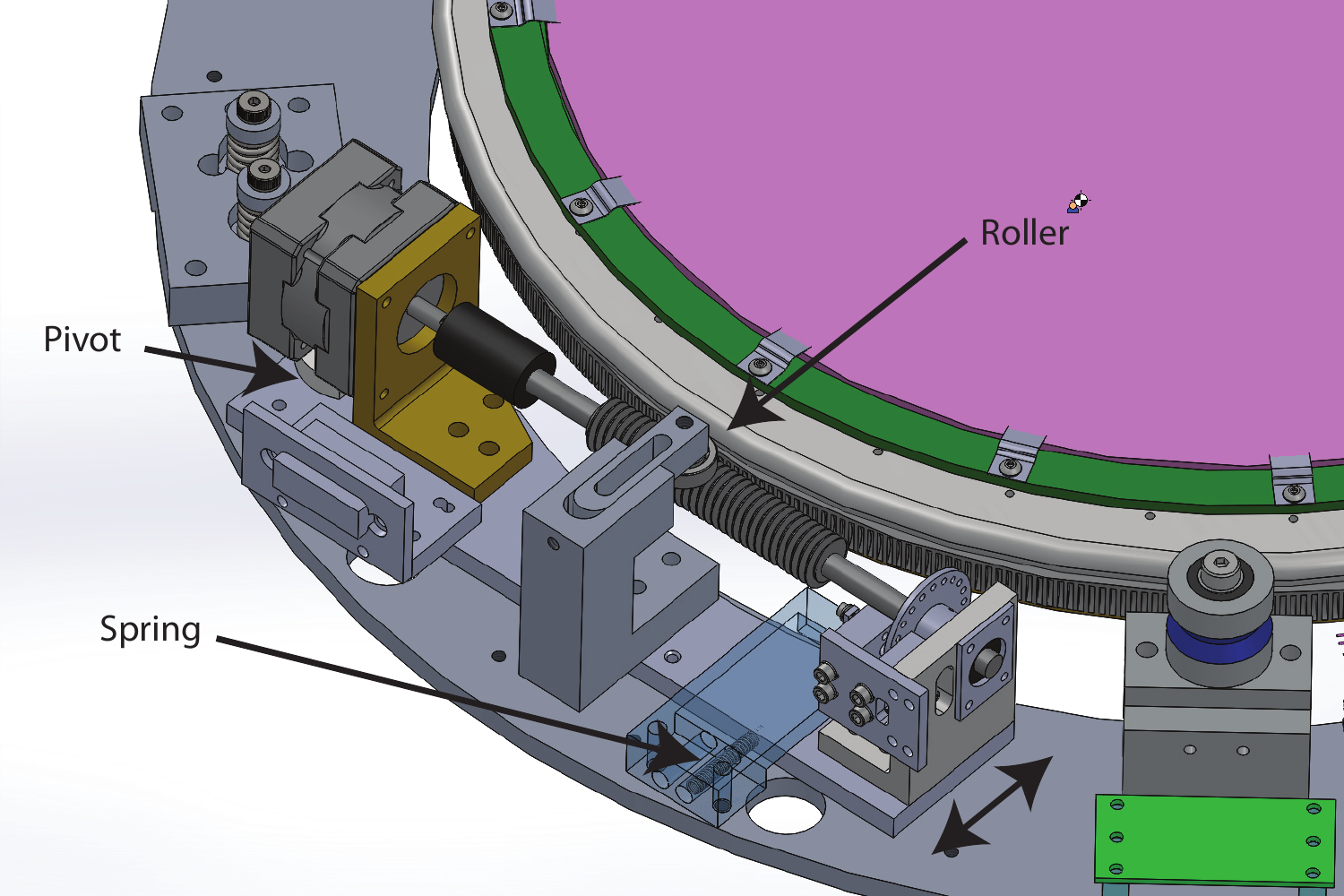}
\caption{Illustration of the kinematic mount and spring loaded worm gear. The left panel shows one of the three kinematic mount points used for a single rotation mechanism. The bolts which attach to the cold plate press on the rotation mechanism rigid mounting plate via springs.  The springs allow sufficient compliance parallel to the plage and a ball bearing to allow the cold plate to distort without distorting the main rotation bearing plate. The right panel shows the spring-loaded worm gear which keeps the worm gear engaged with the main gear. The exact distance between the worm and main gears is maintained with the roller. \label{kinematic_mount_de}}
\end{center}
\end{figure*}

As it arrives from the manufacturer, the motor has an aluminum housing that will shrink onto the new stainless steel ball bearings as they cool, binding them. The coil assembly also shrinks onto the permanent magnet rotor. To prevent this from causing mechanical interference, we increased the clearance in both parts of the motor. To do this, the permanent magnet rotor was mounted in a lathe and its diameter was turned down to 0.859'', a reduction of 0.006''. The motor drive shaft was polished to slip fit inside the ball bearings, allowing for a gentler assembly process that reduces the risk of damage to the bearing or motor shaft. The parts of the aluminum motor housing holding the ball bearing were also mounted in a fixture and enlarged with a reamer to 0.632'', an increase of 0.002''. These modifications are shown in Figure~\ref{motor_mods}. After making all of these modifications, the motor parts were cleaned in acetone and isopropyl alcohol before reassembly.

The drive current for the stepper motor is provided by a Phytron MCC-2 LIN stepper motor controller. This controller uses linear power stages to provide the drive current. This method does not produce RF electrical noise, unlike the more common pulse-width modulation stepper motor drivers. The controllers  do not produce electrical cross talk with the angle encoders, and do not cause electrical problems with the sensitive detectors in the \textsc{Spider} focal plane.

To verify that the modifications were sufficient to allow the motors to perform well when cold, we dunk tested each individual motor in liquid nitrogen. We mounted the motor on an aluminum fixture that could be dunked in a small liquid nitrogen bucket, and used the motor to lift a weight bucket with a pulley system. The torque performance was measured by determining the maximum weight each motor was able to lift. For these tests the motor speed was fixed at 0.5 shaft revolutions per second and a drive current of 0.8 A was used. Each modified motor delivered between 7 and 9 inch-ounces of torque at room temperature, compared with an unmodified motor's performance of 15 inch-ounces. While immersed in liquid nitrogen, the motors delivered between 8 and 11 inch-ounces of torque. This increase upon cooling is expected because the distance between the permanent magnet rotor and coils should decrease upon cooling, increasing the torque.  To remove the water that condensed on the motors as they warmed up after these tests, each motor was disassembled, completely re-cleaned, and assembled again before installation in a rotation mechanism.

Each of the rotation mechanisms were cooled in a pulse tube test cryostat for testing before installation in the \textsc{Spider} flight cryostat. In these measurements, for all of the mechanisms the minimum drive current necessary to turn at 1$^\circ$ per second at $\sim10$~K was measured to be 0.2~A or lower. The minimum current to turn when cold does not change significantly over a speed range from 0.5$^\circ$ per second up to 5$^\circ$ per second. Since the nominal drive current during operation is 0.8~A, this means the mechanisms have a torque safety factor of 4 or more.  The cryogenic heat dissipation of the mechanism was measured by turning two mechanisms simultaneously in the cryostat for 40 minutes with 0.6 A drive current at 1$^\circ$ per second. The cold head of the pulse tube rose to an equilibrium temperature indicating that an additional 0.8~W of loading was present. This load includes heating from the operation of the LEDs in the angle encoders, which is approximately 0.17 W for two mechanisms as shown in Section~\ref{enc_hardware}. At this power level, an individual mechanism turning at 1$^\circ$ per second would boil off only 4 mL of liquid helium in a $22.5^{\circ}$ turn.

\subsection{Floating Kinematic Mount, and Spring-loaded Worm Gear}

We found that the rotation mechanisms, which function well in a small test cryostat, tended to fail when installed in the much larger \textsc{Spider} flight cryostat.  We now believe that the top of the flight cryostat's 4 K tank, onto which the HWP rotation mechanisms are mounted, changes dimensions and distorts slightly when it is cooled from room temperature to 4 K.  Movement of the V-groove bearings relative to one another can disengage the worm drive, which is the failure mode we frequently found.  We implemented two solutions to deal with this problem;  one decouples the rotation mechanism from the 4 K cryostat via a kinematic mount, and the other uses a spring and idler wheel to ensure constant spacing between the two components of the worm gear.

The kinematic mount supports a rigid platform, onto which the HWP, gear mechanism, and encoder components for a single mechanism are all mounted. Each kinematic mount consist of three ball-in-groove interfaces, each with a pair of springs pressing the grooved part (which connects to the rigid platform) onto the ball (which connects to the top of the 4 K tank). The spring force ensures the rigid platform is kept in contact with the 4 K vessel at all times, and was sufficient to survive the shock from launch. The ball-in-groove contact points provide 6 constraints, enough to exactly constrain the necessary 6 degrees of positional freedom, while eliminating stresses or deformations of the rigid platform as the 4 K vessel top plate shrinks or distorts. 

To prevent binding or disengagement between the main and worm gears, a spring loaded idler wheel (contacting the rotor) was implemented to set the clearance of the worm gear. The spring pushes the worm gear against the rotor, while the position of the roller sets the clearance while allowing rotor movement. This means that even in the presence of thermal contraction distortions of the main gear, or small irregularities in the manufacture of the main gear, the exact distance of the worm gear to the main gear will be maintained as the main gear rotates. Figure \ref{kinematic_mount_de} shows a diagram of the kinematic mount and spring-load mechanism design. The fact that all six mechanisms turned well during the flight indicates that the kinematic mount and spring-loaded worm gear were successful.

\section{Angle Encoders}

\subsection{Hardware}\label{enc_hardware}

\begin{figure}
\begin{center}
\includegraphics[width=0.48\textwidth]{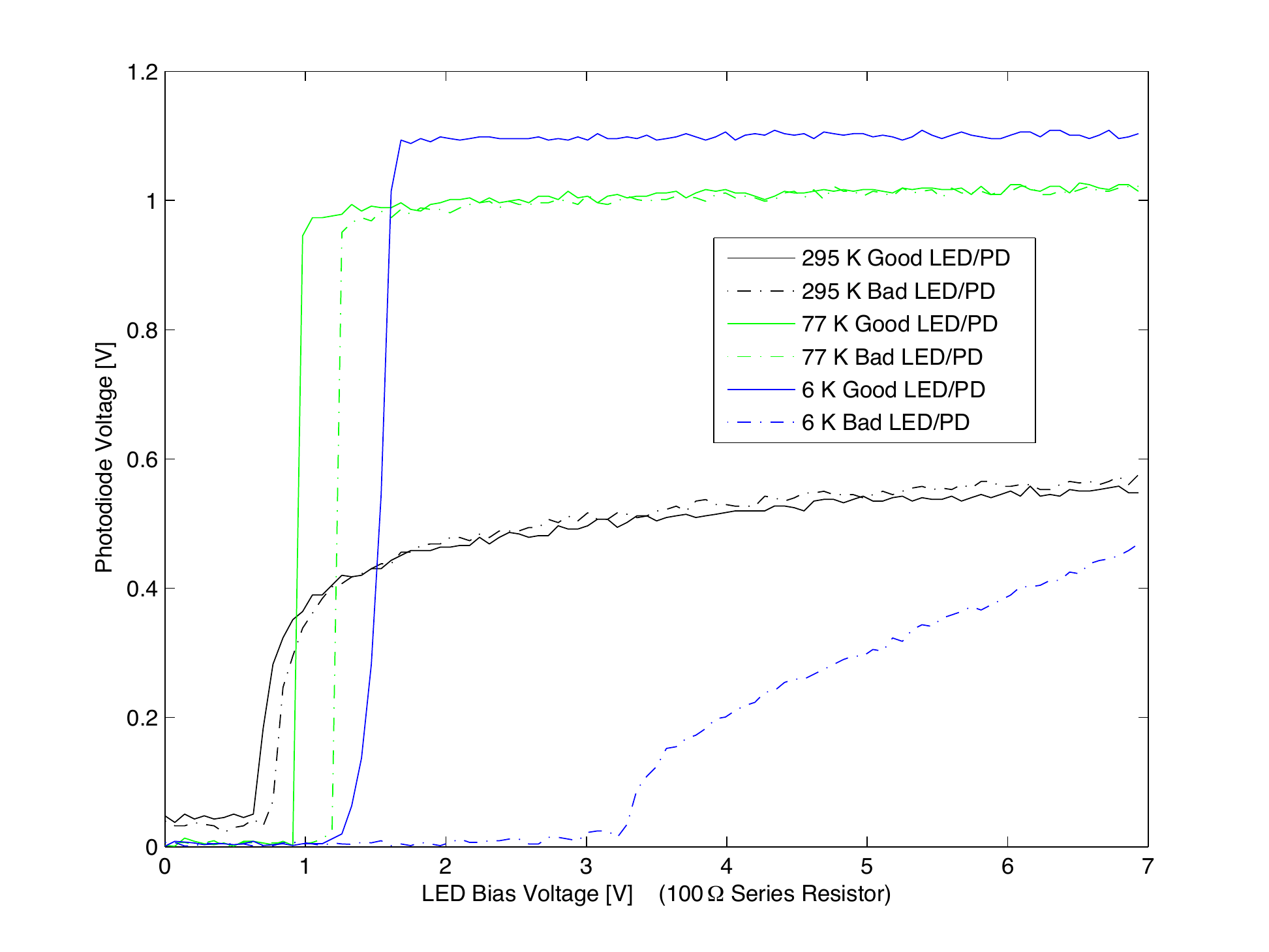}
\caption{Comparison of an operational and defective LED/photodiode pair.  In the operational pair (solid lines), by the time the LED bias voltage exceeds 2 V the photodiode is saturated. The defective pair (dot-dashed lines) operates normally at room and liquid nitrogen temperatures, but at liquid helium temperatures the photodiode fails to saturate  for any of the LED bias voltages. \label{bad_ledpd}}
\end{center}
\end{figure}

We monitor the bearing angle with a precision of $\pm 0.1$$^\circ$. To achieve this, we built a system of several cryogenic optical encoders to measure the rotor bearing angle while it is turning, which allows us to solve for the end position. Each encoder uses a light emitting diode (LED)\footnote{Industrial Fiberoptics IF-E91A, Digi-Key, Thief River Falls, MN www.digikey.com} and a photodiode detector\footnote{Vishay BPV23NF}, operating in a band centered at 940 nm. Although the manufacturers do not guarantee cryogenic performance, these components are inexpensive, readily available, and typically work well at liquid helium temperatures. We screen for units that work well by cooling each LED/photodiode pair to 6~K before choosing pairs to use in a rotation mechanism.  For this screening test, the LED and photodiode are mounted directly facing each other, with the LED DC voltage-biased with a programmable power supply and a 100~$\Omega$ series resistor. A typical dip test result of a functional and defective pair is shown in Figure~\ref{bad_ledpd}. 74 out of 84 tested pairs functioned well at 6~K.

To monitor the rotation of the bearing's stepper motor shaft, we placed an incremental encoder wheel on the motor shaft. The LED shines light towards the photodiode detector, and the beam is alternately blocked and passed as 20 equally-spaced holes rotate through the beam, creating a chopped light signal on the detector.  Each period of the shaft encoder signal corresponds to a main bearing rotation of $(360^{\circ}/463~\mathrm{main~gear~teeth}) \times (1~\mathrm{tooth}/20~\mathrm{shaft~encoder~periods}) = 0.04^{\circ}$.  We use this shaft encoder mainly as a diagnostic.

\begin{figure}
\begin{center}
\subfigure{\includegraphics[width=0.48 \textwidth]{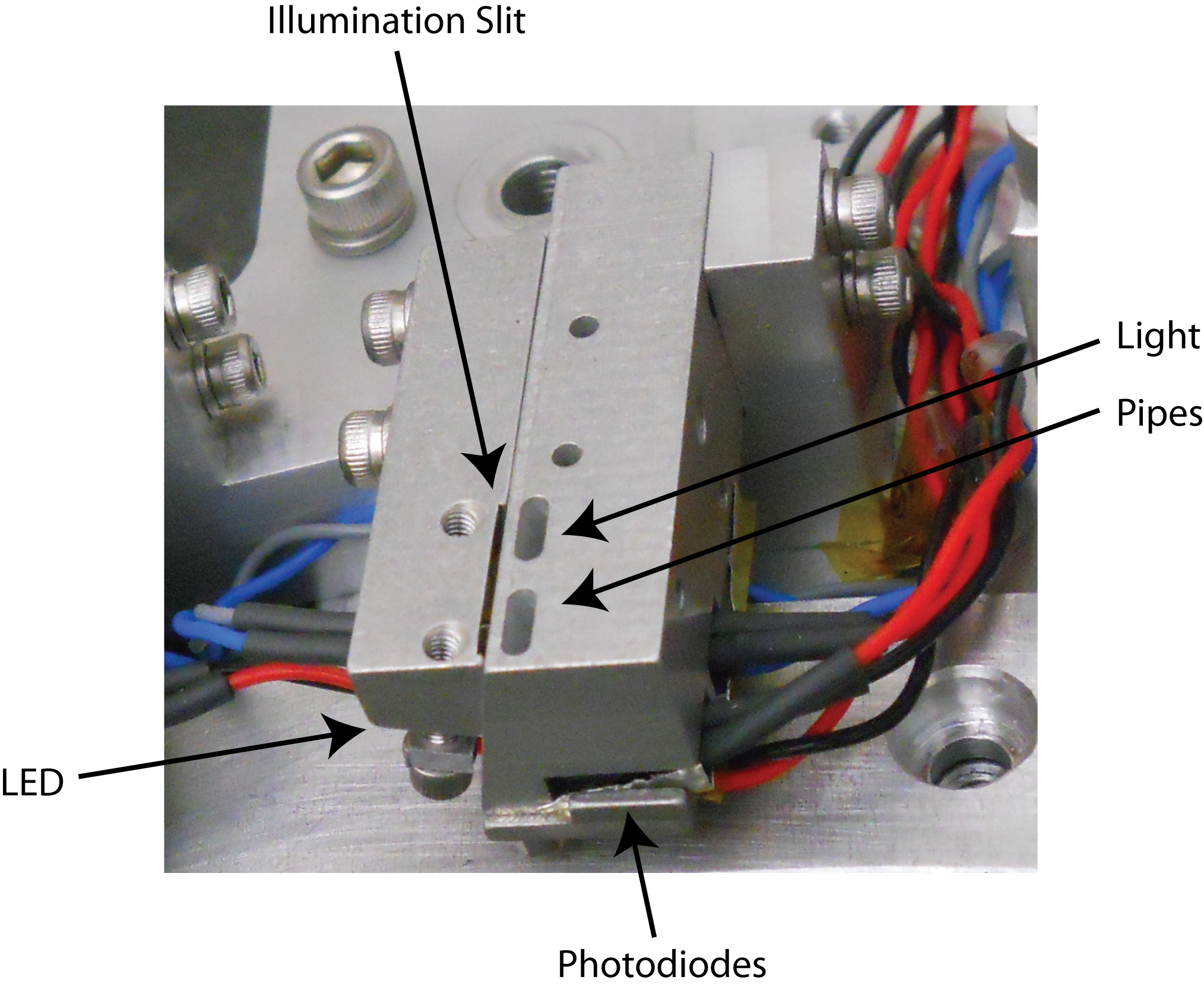}}
\subfigure{\includegraphics[width=0.48 \textwidth]{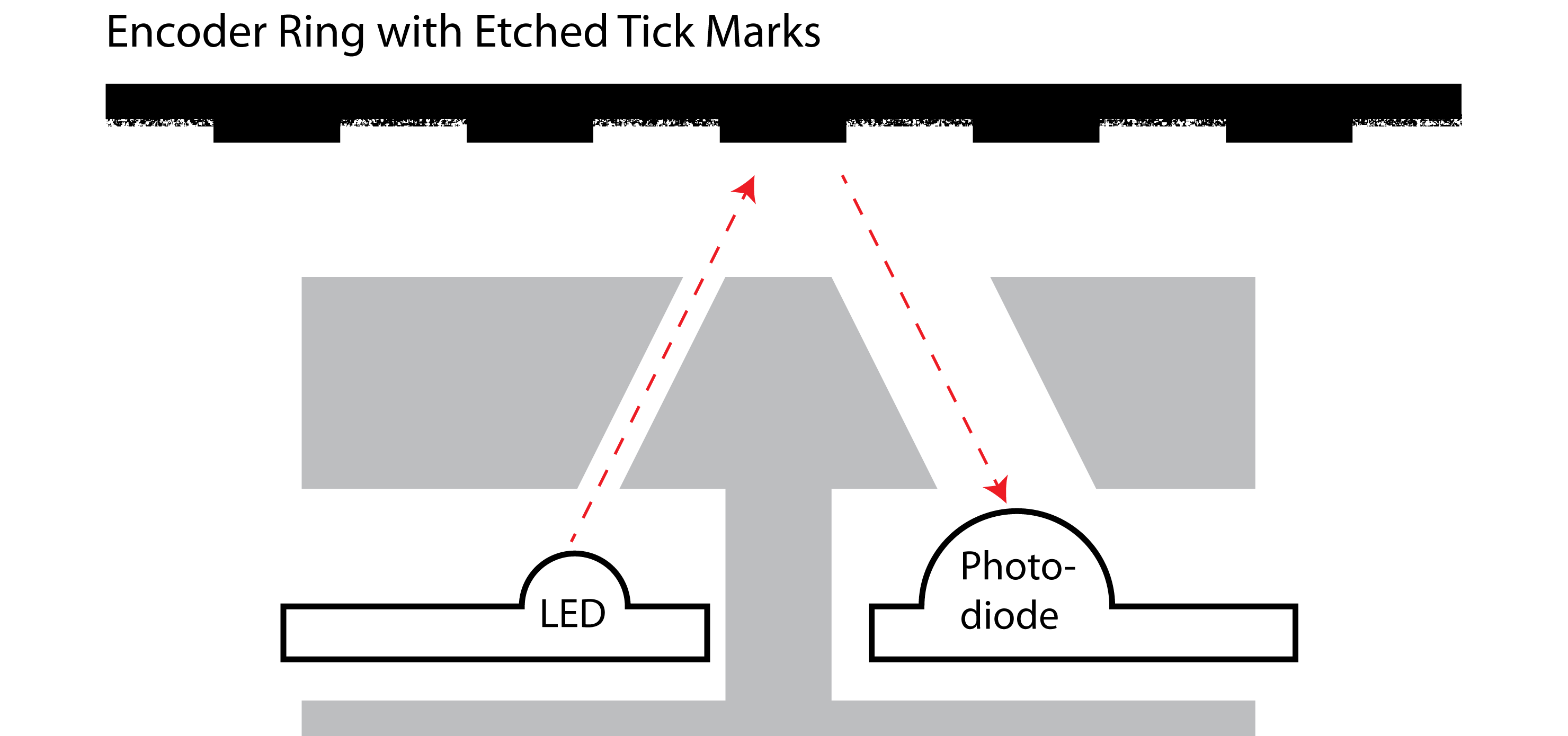}}
\caption{LED and photodiode mount to read out the reflective encoder on the bottom of the main bearing. The LED shines up through the narrow illumination slit and reflects off both tracks of the encoder. One light pipe for each track of the encoder carries light down to the photodiodes for detection.\label{encoder_led_pd_mount}}
\end{center}
\end{figure}

\begin{figure}
\begin{center}
\subfigure{\includegraphics[width=0.48\textwidth]{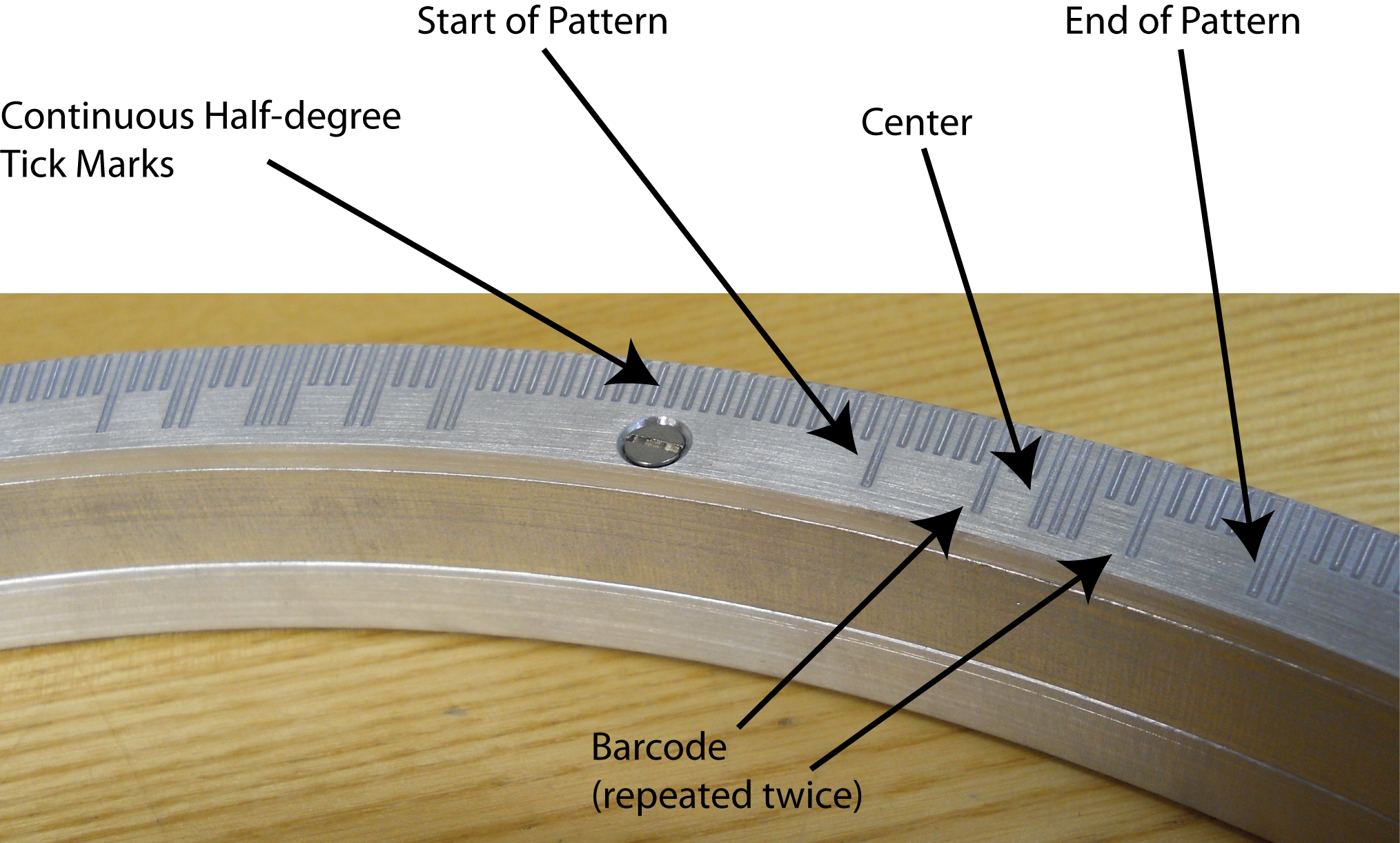}}
\subfigure{\includegraphics[width=0.48\textwidth]{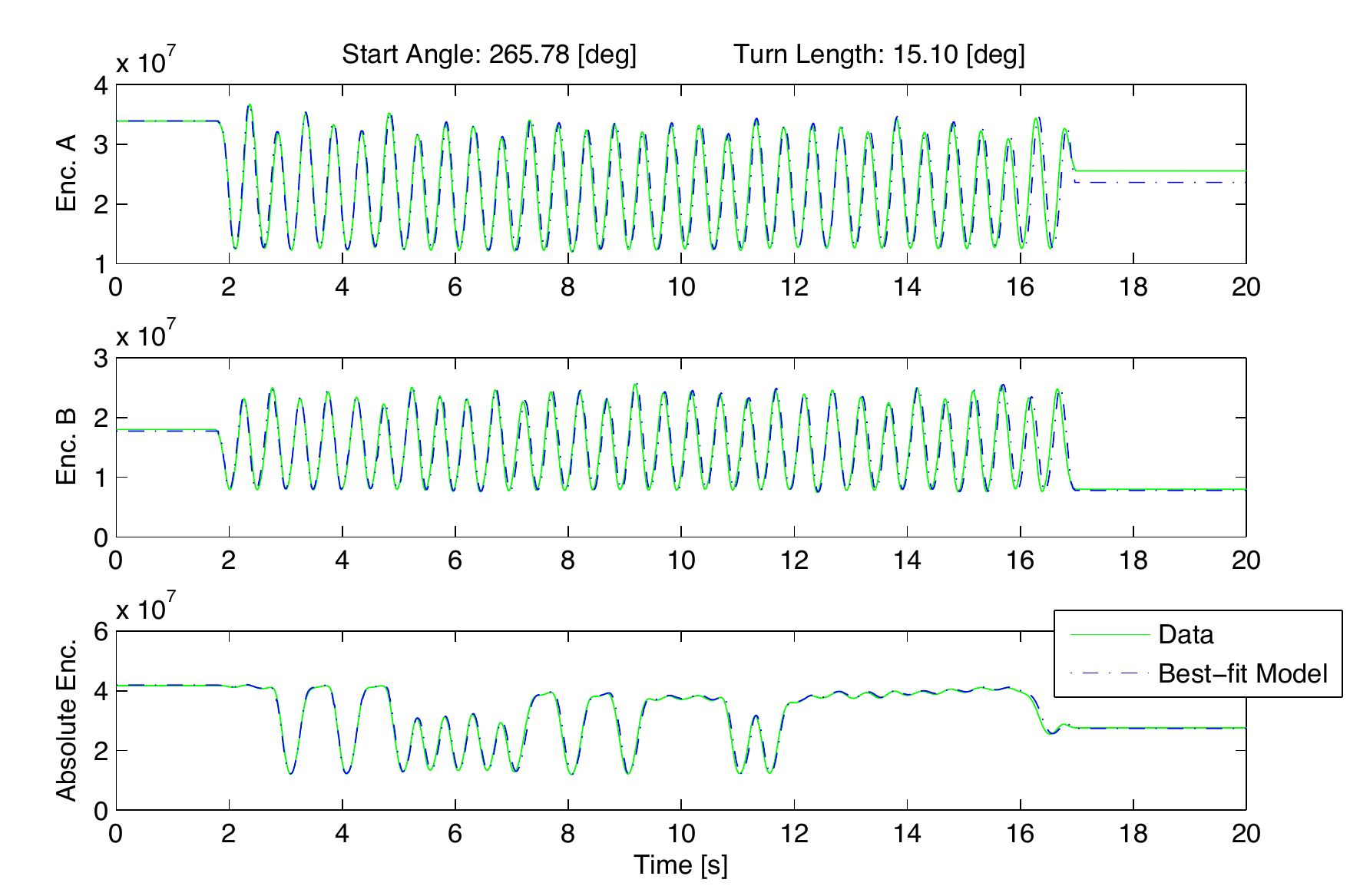}}
\caption{Main encoder. The top panel is a photograph showing both the continuous and barcode tracks of the encoder attached to the main gear. The bottom panels are plots of encoder data (in raw ADC units) from a 15$^\circ$ turn taken with a mechanism operating cold in the \textsc{Spider} flight cryostat. The best-fit model timestream is overplotted, which allows a precise estimate of the start and stop angles. \label{turn_timestream}}
\end{center}
\end{figure}

We also mounted an incremental encoder with absolute reference marks directly on the rotor.  To conserve space, the photodiode views the LED in reflection, as shown in Figure~\ref{encoder_led_pd_mount}. The reflecting surface (attached to the rotor) is patterned by laser etching tickmarks on aluminum. As shown in Figure~\ref{turn_timestream}, the laser-etched tick marks are less reflective than bare aluminum, presumably because of the higher surface roughness of the laser-etched surface. Experimentally, we found that 0.020''-wide ticks were the smallest that were both easily resolvable and yielded high signal-to-noise on the photodiodes with our design. Tick marks were etched\footnote{Laser Design and Services, Willoughby, OH www.ldas-us.com} every $0.5^{\circ}$ on one track around the main bearing. A second track was etched with a unique barcode pattern every $22.5^{\circ}$, shown in Figure~\ref{turn_timestream}. The barcodes allow the absolute start and stop angle of a partial turn that crosses a unique barcode pattern (every turn longer than $22.5^{\circ}$ does this) to be determined solely from that turn's encoder data.  After implementing this design, we found that our two-track scheme is conceptually similar to the Virtual Absolute encoders made by Gurley Precision Instruments. 

We use LED/photodiode pairs to sense the encoder ticks on the main bearing as they pass by. The mounting assembly is shown in Figure~\ref{encoder_led_pd_mount}. The LED shines up through an illumination slit onto both encoder tracks. The slit is sized such that light illuminates one encoder tick at a time. Two photodiode detectors are mounted below two light pipes, with one for each track of the encoder. An additional LED/photodiode pair (not shown in the Figure) also views the continuous track, but is staggered by a quarter-tick, forming a quadrature encoder readout. This makes the angular sensitivity more uniform because as the bearing turns one of the two continuous encoder signals is always transitioning between high and low, and it also indicates the direction of the turn. 

To reduce stray pickup from the stepper motor drive current, the encoders are read out with a lock-in amplifier system. The switching is accomplished using a MOSFET\footnote{Vishay SI2318DS-T1-E3, Digi-Key, Thief River Falls, MN www.digikey.com} in series with each LED driven by a digital logic signal. The LEDs are voltage biased at 6 V with a 100~$\Omega$ series resistor. Since it shines directly onto the photodiode, a 1 k$\Omega$ series resistor is used for the shaft encoder LED to prevent it from saturating the photodiode. Each LED is chopped at 1.5 kHz. When cooled, the voltage drop across this model of LED is approximately 2~V, which implies a calculated power dissipation at 4 K of approximately 80~mW for each main encoder LED, and 8 mW for the shaft encoder LED. The LEDs are chopped at a 50\% duty cycle, which means the total power dissipation for a single mechanism is calculated to be 84 mW.  If all six mechanisms were left running during the entire flight, this heat load would have been too large for the \textsc{Spider} flight cryostat. Since the worm gear prevents any motion even while the motors are powered down, we only powered up the encoder during HWP moves. This means that the encoder LEDs boiled off very little liquid helium during the flight.

The two wires carrying current from a photodiode are connected across a $30~\mathrm{k}\Omega$ resistor at room temperature, and the readout electronics measure the voltage across that resistor. For lab testing we built an analog lock-in circuit. Each channel used an instrumentation amplifier\footnote{Analog Devices AD620} as a preamp for a balanced modulator/demodulator\footnote{Analog Devices AD630} followed by a low-pass active filter\footnote{Texas Instruments UAF42} at 500 Hz. We chose not to deploy this readout system for \textsc{Spider}, since the thermometer readout electronics \cite{benton14} for the flight cryostat have a digital lock-in system that can demodulate the photodiode signals. Biasing the LED creates a 10-100 mV demodulated signal in the photodiode detector with very low noise. The system is readout noise limited with both the analog ($\sim$0.5 mV noise) and digital ($\sim$0.1 mV noise) lock-in electronics.

\subsection{Measuring the Bearing Rotation Angle}

The analog encoder signals are analyzed to precisely estimate the start and stop angle of a turn. To do this, we begin by using data taken earlier from a single long continuous turn to create a template of all the encoder voltages as a function of absolute rotation angle. Either by eye or using automated software, we then use this earlier data to identify which of the barcode patterns went past the encoders during the turn. We then count the continuous encoder ticks to determine the exact start and stop angle relative to the center of that particular barcode pattern. Because the absolute DC level of the continuous encoders is repeatable, we use the DC level of the start and stop to interpolate between ticks to get estimates of the start and stop angle that are much more precise than the $0.25^\circ$ spacing between encoder minima and maxima.

We verified the precision and accuracy of this angle encoding method by checking it on the bench at room temperature against an optical relative encoder\footnote{8225-6000-DQSD Gurley Precision Instruments, Troy, NY www.gpi-encoders.com} connected to the main bearing. The manufacturer certifies that this encoder is accurate to $\pm 0.03^{\circ}$ which is sufficient to verify whether or not the mechanisms meet their design goal of $\pm 0.1^{\circ}$ angle accuracy. We tested four of the seven mechanisms, and they performed equally well within measurement errors. Each half-degree tick in our encoder is $0.50^{\circ} \pm 0.03^{\circ}$ apart (mean $\pm$ RMS) according to the reference encoder. When we commanded a set of 10 $22.5^{\circ}$ turns from the motor controller, the resulting turns from the worst performing mechanism were measured by the reference encoder to be $22.56^{\circ} \pm 0.06^{\circ}$ long. The turn lengths estimated by our encoder differed from the reference encoder by $0.01^{\circ} \pm 0.05^{\circ}$. We also commanded a set of 14 forward and reverse turns of lengths varying from $22.5^{\circ}$ to $157.5^{\circ}$. In the worst performing mechanism the reference encoder estimates differed from our encoder by $0.00^{\circ} \pm 0.10^{\circ}$. Our encoder measured that in each pair of forward and reverse turns, the mechanism's stop angle differed from its starting angle by $0.01^{\circ} \pm 0.05^{\circ}$. This set of tests verifies that on the bench, we can reliably control and measure the absolute angle of the bearing to better than $\pm 0.1^{\circ}$. Upon cooling, the encoder voltage template does not change appreciably, only differing by an overall gain and DC offset caused by thermal effects in the LED and photodiode. We interpret this as evidence that, when cooled inside a cryostat, the encoders perform equally well as was measured at room temperature on the bench.

\section{Conclusions}

We constructed, tested, and successfully deployed six cryogenic rotation mechanisms for the half-wave plates in the \textsc{Spider} instrument. They have a large 305 mm clear aperture but use small bearings, reducing the impact of differential thermal contraction. This means that this design should be scalable to larger sizes. The mechanisms meet their performance specifications of precise angle control, secure mounting of the HWP/AR stack, low power dissipation, and mechanical durability. The stepper motor driving the mechanism is available commercially and after modifications it functions well at cryogenic temperatures. The angle encoder hardware yields precise measurements of the absolute bearing angle while taking up only a small amount of radial space. The mechanisms performed well during \textsc{Spider}'s first Antarctic flight. More than 138 successful move commands were executed by the six HWP mechanisms during flight, and at all times the HWPs were within a half-degree of their commanded angle. Because the motors and encoders only needed to be powered twice a day while stepping between angles, the mechanisms boiled off very little liquid helium during the flight. The mean and RMS deviation from the intended angle was 0.05$^\circ$ +/- 0.15$^\circ$ which was much better than the precision of 1$^\circ$ required for the science. Based on this success, we hope to use a similar design again in the second flight planned for 2017-2018.

\section{Acknowledgements}

\textsc{Spider} is supported in the U.S. by National Aeronautics and Space Administration under Grant No. NNX07AL64G and NNX12AE95G issued through the Science Mission Directorate, with support for ASR from NESSF NNX10AM55H, and by the National Science Foundation through PLR-1043515. Logistical support for the Antarctic deployment and operations was provided by the NSF through the U.S. Antarctic Program. The collaboration is grateful for the generous support of the David and Lucile Packard Foundation, which has been crucial to the success of the project.

Support in Canada is provided by the National Sciences and Engineering Council and the Canadian Space Agency.

\bibliographystyle{unsrt}
\bibliography{bibliography}

\end{document}